\documentclass[10pt, a4paper]{article}
\usepackage{wds10}

\usepackage[utf8]{inputenc}

\usepackage{amsmath}

\usepackage{amsfonts}
\usepackage{mathrsfs}

\usepackage{units}

\usepackage[square]{natbib}

\usepackage{placeins}

\usepackage{verbatim}

\usepackage{accents}

\usepackage[absolute, overlay]{textpos}
\newcommand{\mrm}{\mathrm}

\makeatletter

\let\old@dmathbeg\[
\let\old@dmathend\]

\newcommand{\rovnec}[1]{\old@dmathbeg#1\old@dmathend}
\newcommand{\rovcis}[2]{\begin{equation}#1\label{#2}\end{equation}}
\newcommand{\drovcis}[2]{\begin{equation}\begin{split}#1\end{split}\label{#2}\end{equation}}
\newcommand{\drovnec}[1]{\begin{equation*}\begin{split}#1\end{split}\end{equation*}}
\newcommand{\provcis}[1]{\begin{align}#1\end{align}}
\newcommand{\provnec}[1]{\begin{align*}#1\end{align*}}

\newcommand{\rov}{\@ifstar\rovnec\rovcis}
\newcommand{\drov}{\@ifstar\drovnec\drovcis}
\newcommand{\prov}{\@ifstar\provnec\provcis}

\newcommand{\vast}{\bBigg@{4}}
\newcommand{\Vast}{\bBigg@{5}}

\makeatother

\newcommand{\vld}{\underaccent{\tilde}}

\DeclareMathOperator{\diffbold}{\mathbf{d}}
\newcommand{\bd}{\diffbold\!}

\newcommand{\msc}{\mathscr}
\newcommand{\mbs}{\boldsymbol}

\newcommand{\iDelta}{{\mit\Delta}}
\newcommand{\iSigma}{{\mit\Sigma}}
\newcommand{\iLambda}{{\mit\Lambda}}

\renewcommand{\[}{\left[}
\renewcommand{\]}{\right]}

\newcommand{\f}{\!\left}

\newcommand{\pder}[3][]{\frac{\partial^{#1}#2}{\partial{#3}^{#1}}}

\newcommand{\zrov}{{}\\{}}
\newcommand{\nzrov}{{}\nonumber\\{}}
\newcommand{\zrovn}{{}\\\nonumber{}}
\newcommand{\res}[2]{\left.#1\right|_{#2}}
\newcommand{\lbl}{\label}
\newcommand{\rvt}{\ .}
\newcommand{\rvc}{\ ,}
\newcommand{\qt}[1]{``#1''}

\renewcommand{\(}{\left(}
\renewcommand{\)}{\right)}

\lefthead{Hejda and BIČÁK}
\righthead{Black Holes and Magnetic Fields}

\begin{document}

\begin{textblock}{165.5}(18.5,9.5)
{\small\it\fontfamily{ppl}\selectfont
 WDS'14 Proceedings of Contributed Papers --- Physics, 48–55, 2014.
\hspace*{\fill} ISBN 978-80-7378-276-4 \copyright{} MATFYZPRESS}
\end{textblock}

\begin{textblock}{}(96.2,284.8)
{\fontfamily{ppl}\selectfont
 48}
\end{textblock}

\title{Black Holes and Magnetic Fields}

\author{Filip Hejda and Jiří Bičák}

\affil{Institute of Theoretical Physics, Charles University, Faculty  of Mathematics and Physics, Prague, Czech Republic}

\begin{abstract}
We briefly summarise the basic properties of spacetimes representing rotating, charged black holes in strong axisymmetric magnetic fields. We concentrate on extremal cases, for which the horizon surface gravity vanishes. We investigate their properties by finding simpler spacetimes that exhibit their geometries near degenerate horizons. Employing the simpler geometries obtained by near-horizon limiting description we analyse the Meissner effect of magnetic field expulsion from extremal black holes.
\end{abstract}

\begin{article}

\section{Charged, rotating black holes in magnetic fields}

A simple solution to the Einstein--Maxwell system of equations, which is often referred to as a \qt{magnetic universe}, was studied by \cite{Melvin64}. The metric can be expressed in cylindrical coordinates ($R=r\sin\vartheta$, $z=r\cos\vartheta$) as
\rov{\mbs g=\(1+\frac{1}{4}B^2R^2\)^2\(-\bd t^2+\bd R^2+\bd z^2\)+\frac{R^2}{\(1+\frac{1}{4}B^2R^2\)^2}\bd\varphi^2\rvc}{mmu}
accompanied with the magnetic field, characterised by parameter $B$, with only nonzero component of the electromagnetic potential
\rov{A_\varphi=\frac{\frac{1}{2}BR^2}{1+\frac{1}{4}B^2R^2}\rvt}{afimmu}

Solution generating technique known as Harrison transformation applied on a \qt{seed} Minkowski spacetime yields this Melvin magnetic universe. Therefore, as one may expect, if we apply the Harrison transformation to asymptotically flat black hole solutions, the results are black holes immersed in magnetic universes. This possibility was studied by \cite{Ernst76}.

\subsection{Stationary Ernst solution}

A useful example of a Harrison-transformed spacetime is the stationary Ernst solution
\drov{\mbs g=\left|\iLambda\right|^2\[-\(1-\frac{2M}{r}+\frac{Q^2}{r^2}\)\bd t^2+\frac{1}{1-\frac{2M}{r}+\frac{Q^2}{r^2}}\bd r^2+r^2\bd\vartheta^2\]+\zrov+\frac{r^2\sin^2\vartheta}{\left|\iLambda\right|^2}\(\bd\varphi-\omega\bd t\)^2\rvc}{syE}
which represents a Reissner--Nordström black hole with mass parameter $M$ and charge parameter $Q$ in an external magnetic field. The influence of the magnetic field on the geometry is
 expressed by a complex function $\iLambda$:
\rov{\iLambda=1+\frac{1}{4}B^2\(r^2\sin^2\vartheta+Q^2\cos^2\vartheta\)-\mrm iBQ\cos\vartheta\rvt}{lamsyE}
The electromagnetic field is more complicated than just a superposition of the electrostatic field of the black hole and some simple external field. This arises from the non-linear nature of the Einstein--Maxwell system. However, the actual field has the same crucial property as that simple superposition would have: a non zero angular momentum. This induces frame dragging with dragging potential
\rov{\omega=-\frac{2BQ}{r}+B^3Qr+\frac{B^3Q^3}{2r}-\frac{B^3Q}{2r}\(r^2-2Mr+Q^2\)\sin^2\vartheta\rvt}{omsyE}

One can get azimuthal component of the electromagnetic potential also by means of Harrison transformation (\cite{Pope})
\rov{A_\varphi^{\(1\)}=\frac{1}{B\left|\iLambda\right|^2}\[2\Re\iLambda\(\Re\iLambda-1\)+\(\Im\iLambda\)^2\]\rvt}{}
However, the given gauge is not very convenient because $A_\varphi^{\(1\)}$ does not vanish on the axis. This can be fixed by subtracting the value of \qt{raw} $A_\varphi^{\(1\)}$ for $\vartheta=0$. In this way we obtain
\rov{A_\varphi=\frac{1}{B\left|\iLambda\right|^2}\[\frac{\(2+\frac{3}{2}B^2Q^2\)\(\Re\iLambda\)^2+\(1-\frac{1}{16}B^4Q^4\)\(\Im\iLambda\)^2}{1+\frac{3}{2}B^2Q^2+\frac{1}{16}B^4Q^4}-2\Re\iLambda\]\rvt}{AfisyE}
This unique gauge is important for the study of particle motion in the Hamiltonian formalism, since there the electromagnetic potential becomes a physically relevant quantity.

Another important quantity is the tetrad component describing radial electric field strength
\drov{F_{(r)(t)}={}&\frac{1}{\left|\iLambda\right|^4}\Bigg\{\[\(\Re\iLambda\)^2-\(\Im\iLambda\)^2\]\(2-\Re\iLambda\)\frac{Q}{r^2}+\zrov&+\frac{B}{2}\(1-\frac{Q^2}{r^2}\)\Im\iLambda\[\(\Re\iLambda\)^2-\(\Im\iLambda\)^2-4\Re\iLambda\]\cos\vartheta\Bigg\}\rvc}{F20syE}
which directly enters some relations that we will discus further. (It can also be used to determine the form of the component $A_t$.)

\subsection{Ernst--Wild solution and a general MKN black hole}

\begin{textblock}{}(96.2,284.8)
{\fontfamily{ppl}\selectfont
 49}
\end{textblock}
\vspace{0.25em}

\cite{ErnstWild} unveiled another solution that is obtained by applying the Harrison transformation to the Kerr (uncharged) \qt{seed} metric with rotation parameter $a$. The resulting metric
\rov{\mbs{g}=\left|\iLambda\right|^2\iSigma\[-\frac{\iDelta}{\msc{A}}\bd t^2+\frac{\bd r^2}{\iDelta}+\bd\vartheta^2\]+\frac{\msc A}{\iSigma\left|\iLambda\right|^2}\sin^2\vartheta\(\bd\varphi-\omega\bd t\)^2}{ewmkn}
contains functions $\iDelta, \iSigma, \msc A$ in the form
\prov{\iSigma&=r^2+a^2\cos^2\vartheta\rvc&\msc A&=\(r^2+a^2\)^2-\iDelta a^2\sin^2\vartheta\rvc&\iDelta&=r^2-2Mr+a^2\rvc}
 from the \qt{seed} metric and a new (complex) metric function $\iLambda$, which has the form
\rov{\iLambda=1+\frac{1}{4}B^2\frac{\msc A}{\iSigma}\sin^2\vartheta-\frac{\mrm{i}}{2}B^2Ma\cos\vartheta\(3-\cos^2\vartheta+\frac{a^2}{\iSigma}\sin^4\vartheta\)\rvt}{lamew}
The new dragging potential is
\drov{\omega=\frac{a}{r^2+a^2}\Bigg\{\(1-B^4M^2a^2\)-\iDelta\Bigg[\frac{\iSigma}{\msc A}+\frac{B^4}{16}\Bigg(-8Mr\cos^2\vartheta\(3-\cos^2\vartheta\)-\zrov-6Mr\sin^4\vartheta+\frac{2Ma^2\sin^6\vartheta}{\msc A}\[r\(r^2+a^2\)+2Ma^2\]+\zrov+\frac{4M^2a^2\cos^2\vartheta}{\msc A}\[\(r^2+a^2\)\(3-\cos^2\vartheta\)^2-4a^2\sin^2\vartheta\]\Bigg)\Bigg]\Bigg\}\rvt}{omew}
The components of the field strength tensor are quite complicated.

The general MKN (magnetised Kerr--Newman) black hole has the metric \eqref{ewmkn} with $\iDelta$ from the Kerr--Newman metric, i.e. $\iDelta=r^2-2Mr+a^2+Q^2$, and with much more complicated $\iLambda$:
\drov{\iLambda=1&+\frac{1}{4}B^2\(\frac{\msc A+a^2Q^2\(1+\cos^2\vartheta\)}{\iSigma}\sin^2\vartheta+Q^2\cos^2\vartheta\)+\zrov&+\frac{BQ}{\iSigma}\[ar\sin^2\vartheta-\mrm{i}\(r^2+a^2\)\cos\vartheta\]-\zrov&-\frac{\mrm{i}}{2}B^2a\cos\vartheta\[M\(3-\cos^2\vartheta\)+\frac{Ma^2\sin^2\vartheta-Q^2r}{\iSigma}\sin^2\vartheta\]\rvt}{lammkn}
The most complete and comprehensive picture of the properties of a general MKN black hole is given by \cite{Pope}\footnote{The authors even performed the integration of the electromagnetic potential. However, the original version of the paper published on arXiv contained a number of misprints. We were in contact with the authors, who now made corrections.}. The expression for the dragging potential is given by equations (B.8)--(B.9) in \cite{Pope}, whereas the components of electromagnetic potential are described by formulae (B.15)--(B.18).

\subsection{Asymptotics of magnetised black holes}

The asymptotic properties of the magnetised black hole spacetimes can be so overwhelming for large values of dimensionless $BM$ that there is no approximately flat region in such a spacetime. In that case the astrophysical meaning is doubtful. The conditions for existence of an approximately flat region have been discussed by \cite{BiJa85}. The region must be well outside the horizon, but it must also hold that $\left|\iLambda\right|^2$ is approximately unity in that region. These two requirements are satisfied when $r$ satisfies inequality $r_+\ll r\ll \nicefrac{1}{B}$. If we consider an extremal MKN black hole, i.e. $r_+=r_0=M$, at radii about $10M$ we need $BM$ about $0{.}01$ or less in order for the spacetime to be approximately flat in that region. On the other hand, it is clear that for $BM>0{.}1$ any magnetised black hole spacetime cannot contain even very roughly flat region. This is confirmed by the appearance of embedding diagrams by \cite{StuchHle99}.

\section{The near-horizon description of extremal configurations}

\begin{textblock}{}(96.2,284.8)
{\fontfamily{ppl}\selectfont
 50}
\end{textblock}
\vspace{0.25em}

The near-horizon description of the Kerr--Newman black hole was first considered by \cite{Carter73}. More precise formulation can be found in a more recent paper by \cite{BardHorow}. We will apply their approach to black holes in magnetic fields. Let us emphasise that the near-horizon description assumes extremal black hole configurations, i.e. in the case of MKN we impose condition $M^2=Q^2+a^2$.

\subsection{The metric in the extremal case}

The general metric of an extremal black hole can be written in the form
\rov{\mbs g=-\(r-r_0\)^2\tilde N^2\bd t^2+g_{\varphi\varphi}\(\bd\varphi-\omega\bd t\)^2+\frac{\tilde g_{rr}}{\(r-r_0\)^2}\bd r^2+g_{\vartheta\vartheta}\bd\vartheta^2\rvc}{axst2}
where the degenerate horizon is located at $r=r_0$ and $\tilde N, \tilde g_{rr}$ are regular and non-vanishing there.

To \qt{drag} the coordinates to the \qt{near-horizon} region we introduce new coordinates $\tau, \chi$ by relations
\prov{r&=r_0+p\chi\rvc&t&=\frac{\tau}{p}\rvt\label{rchittau}}
The transformation depends on a limiting parameter $p$, and for any finite nonzero value of this parameter the new coordinates will cover entire spacetime up to \qt{standard}  spatial infinity. However, in extremal black hole spacetimes, there is yet another infinity, since proper radial distance between two points along $t=\mrm{const.}$ diverges, if one of the points approaches $r_0$. When the parameter $p$ in transformation \eqref{rchittau} converges to zero, we get a new metric which describes the infinite region (\qt{throat}) around $r=r_0$. Standard spatial infinity is lost in this near-horizon limiting description.

Let us specify in more detail, what happens with the expression in the metric that contains the dragging potential. We should first expand the dragging potential around its value $\omega_\mrm{H}$ at the horizon
\rov{\omega\doteq\omega_\mrm{H}+\res{\pder{\omega}{r}}{r_0}\(r-r_0\)=\omega_\mrm{H}+\res{\pder{\omega}{r}}{r_0}p\chi\rvc}{omnhexp}
so that
\rov{\bd\varphi-\omega\bd t\doteq\bd\varphi-\(\omega_\mrm{H}+\res{\pder{\omega}{r}}{r_0}p\chi\)\frac{\bd\tau}{p}=\bd\varphi-\frac{\omega_\mrm{H}}{p}\bd\tau-\res{\pder{\omega}{r}}{r_0}\chi\bd\tau\rvt}{dragnhexp}
Recalling the rigidity theorem (see \cite{FrolNov}), which guarantees that $\omega_\mrm{H}$ is a constant, we can write the transformation from $\varphi$ to a \qt{rewinded angle} $\psi$
\rov{\varphi=\psi+\frac{\omega_\mrm{H}}{p}\tau\rvt}{phipsi}
Then we have
\rov{\bd\varphi-\omega\bd t\doteq\bd\psi-\res{\pder{\omega}{r}}{r_0}\chi\bd\tau\rvc}{}
and we see that the new dragging potential after the limit is just the first order of Taylor expansion of the original one.
Note that \eqref{phipsi} is merely a transformation to rotating coordinates, which can be understood as a \qt{gauge-fixing} of the integration constant of the dragging potential. The usual choice of vanishing $\omega$ at spatial infinity becomes undesirable when we linearise it around the horizon and we must turn to the gauge corotating with the horizon.

If we apply the \qt{recipe} to the Kerr--Newman solution, the resulting metric is (cf. \cite{Carter73}, \cite{BardHorow})
\drov{\mbs g=\[Q^2+a^2\(1+\cos^2\vartheta\)\]\(-\frac{\chi^2}{\(Q^2+2a^2\)^2}\bd\tau^2+\frac{\bd\chi^2}{\chi^2}+\bd\vartheta^2\)+\zrov+\frac{\(Q^2+2a^2\)^2}{Q^2+a^2\(1+\cos^2\vartheta\)}\sin^2\vartheta\(\bd\psi+\frac{2a\sqrt{Q^2+a^2}\chi}{\(Q^2+2a^2\)^2}\bd\tau\)^2\rvt}{knnh}

\subsection{Electromagnetic field}

\begin{textblock}{}(96.2,284.8)
{\fontfamily{ppl}\selectfont
 51}
\end{textblock}
\vspace{0.25em}

Jacobi's matrices that will transform components of tensors into the coordinates $\tau, \chi, \psi$ will contain singular expressions. Therefore not every tensor quantity one can think of will be expressible in those coordinates in the limit $p\to0$.

We now turn to the electromagnetic field and expand it to the linear order near the degenerate horizon:
\drov{\mbs A&=A_t\bd t+A_\varphi\bd\varphi\doteq\zrov&\doteq\(\res{A_t}{r_0}+\res{\pder{A_t}{r}}{r_0}p\chi\)\frac{\bd\tau}{p}+\(\res{A_\varphi}{r_0}+\res{\pder{A_\varphi}{r}}{r_0}p\chi\)\(\bd\psi+\frac{\omega_\mrm{H}}{p}\bd\tau\)\doteq\zrov&\doteq\(\res{A_t}{r_0}+\omega_\mrm{H}\res{A_\varphi}{r_0}\)\frac{\bd\tau}{p}+\(\res{\pder{A_t}{r}}{r_0}+\omega_\mrm{H}\res{\pder{A_\varphi}{r}}{r_0}\)\chi\bd\tau+\res{A_\varphi}{r_0}\bd\psi=\zrov&=-\frac{\phi_\mrm{H}}{p}\bd\tau+\(\res{\pder{A_t}{r}}{r_0}+\omega_\mrm{H}\res{\pder{A_\varphi}{r}}{r_0}\)\chi\bd\tau+\res{A_\varphi}{r_0}\bd\psi\rvt}{emnhlim}
Since $\phi_\mrm{H}=\mrm{const.}$, the singular term can be subtracted from the final \qt{limitng} potential and we may consider it to be a gauge constant.

For the particular case of the Kerr--Newman solution the near-horizon electromagnetic potential stripped of the singular gauge constant reads\footnote{This field seems to be more appropriate than the one given in \cite{Carter73} --- see \cite{dipl}, Appendix B for discussion --- since Carter's expression does not satisfy Maxwell equations.}
\rov{\mbs A=\frac{Q}{Q^2+a^2\(1+\cos^2\vartheta\)}\(\frac{Q^2+a^2\sin^2\vartheta}{Q^2+2a^2}\chi\bd\tau+a\sqrt{Q^2+a^2}\sin^2\vartheta\bd\psi\)\rvt}{aknnh}

\subsection{Stronger corollaries of rigidity theorems}

It is worth to note that
\rov{F_{(r)(t)}=\frac{1}{N\sqrt{g_{rr}}}\(\pder{A_t}{r}+\omega\pder{A_\varphi}{r}\)\rvt}{}
The $F_{(r)(t)}$ component can be obtained directly from Harrison transformation just by algebraic manipulation, so that it is one of the least laborious quantities for magnetic universes. Recalling \eqref{emnhlim} we can express
\rov{A_\tau=\res{\(\tilde N\sqrt{\tilde g_{rr}}F_{(r)(t)}\)}{r_0}\chi\rvt}{}

It can be shown (\cite{dipl}, Appendix A) that on a degenerate horizon of any extremal MKN black hole it holds that
\prov{\tilde\omega&\equiv\res{\pder{\omega}{r}}{r=r_0}=\mrm{const.}\rvc&\tilde\phi&\equiv\res{\pder{\phi}{r}}{r=r_0}=\mrm{const.}}
We can write for the near-horizon limiting quantities of all the models that we consider
\prov{\omega&=\tilde\omega\chi\rvc&\phi&=\tilde\phi\chi\rvc&A_\tau&=\tilde A_\tau\f(\vartheta\)\chi\rvt\lbl{nhtiqu}}
From the definition of the generalised electromagnetic potential we can express
\prov{\phi&=-A_\tau-\omega A_\psi\rvc&A_\psi\f(\vartheta\)&=\frac{1}{\tilde\omega}\(-\tilde\phi-\tilde A_\tau\f(\vartheta\)\)\rvt}
Since we demand the electromagnetic potential to be continuous, the azimuthal component must vanish at the axis, so
\prov{\phi&=-\res{A_\tau}{\vartheta=0}\rvc&A_\psi\f(\vartheta\)&=\frac{1}{\tilde\omega}\(\tilde A_\tau\f(0\)-\tilde A_\tau\f(\vartheta\)\)\rvt}

Hence the knowledge of $F_{(r)(t)}$ in the original spacetime is sufficient to determine the whole potential of electromagnetic field in the near-horizon spacetime.

\subsection{Near-horizon description of stationary Ernst solution}

\lbl{odd:nhsye}

Let us turn to an extremal stationary Ernst spacetime. In the cases of magnetic universes, the near-horizon description will be advantageous also because we will get rid of the ill, non-flat, asymptotics of the solutions. Applying the recipe described above to metric \eqref{syE}, we get
\drov{\mbs g=\[\(1+\frac{1}{4}B^2Q^2\)^2+B^2Q^2\cos^2\vartheta\]\(-\frac{\chi^2}{Q^2}\bd\tau^2+\frac{Q^2}{\chi^2}\bd\chi^2+Q^2\bd\vartheta^2\)+\zrov+\frac{Q^2\sin^2\vartheta}{\(1+\frac{1}{4}B^2Q^2\)^2+B^2Q^2\cos^2\vartheta}\[\bd\psi-\frac{2B}{Q}\(1+\frac{1}{4}B^2Q^2\)\chi\bd\tau\]^2\rvc}{syEnh}
which for $B=0$ turns to \eqref{knnh} with $a=0$, i.e. to the Robinson--Bertotti solution. The components of the electromagnetic potential can be evaluated directly from $F_{(r)(t)}$ component \eqref{F20syE}:
\prov{A_\tau&=\frac{\chi}{Q}\frac{\(1+\frac{1}{4}B^2Q^2\)^2-B^2Q^2\cos^2\vartheta}{\(1+\frac{1}{4}B^2Q^2\)^2+B^2Q^2\cos^2\vartheta}\(1-\frac{1}{4}B^2Q^2\)\rvc\\A_\psi&=\frac{-BQ^2}{1+\frac{3}{2}B^2Q^2+\frac{1}{16}B^4Q^4}\frac{\(1-\frac{1}{16}B^4Q^4\)\sin^2\vartheta}{\(1+\frac{1}{4}B^2Q^2\)^2+B^2Q^2\cos^2\vartheta}\rvt}
The quantities given above satisfy the full Einstein--Maxwell system. Furthermore we have found that this simple solution is contained in a richer discussion of generalisations of Reissner--Nordström and Robinson--Bertotti solutions, which was carried out by \cite{DiazBaez86}. They also discuss the possibility of a near-horizon limiting process.

\subsection{Near-horizon description of Ernst--Wild solution}

\begin{textblock}{}(96.2,284.8)
{\fontfamily{ppl}\selectfont
 52}
\end{textblock}
\vspace{0.25em}

Having examined one possible generalisation, we may now proceed to somewhat \qt{fearsome} case of the Ernst--Wild solution. Applying the near-horizon limit to metric \eqref{ewmkn}, one gets
\drov{\mbs g=\[\(1+B^4a^4\)\(1+\cos^2\vartheta\)+2B^2a^2\sin^2\vartheta\]\(-\frac{\chi^2}{4a^2}\bd\tau^2+\frac{a^2}{\chi^2}\bd\chi^2+a^2\bd\vartheta^2\)+\zrov+\frac{4a^2\sin^2\vartheta}{\(1+B^4a^4\)\(1+\cos^2\vartheta\)+2B^2a^2\sin^2\vartheta}\[\bd\psi+\frac{\chi}{2a^2}\(1-B^4a^4\)\bd\tau\]^2\rvt}{ewnh}
Note that the dragging potential \eqref{omew} vastly simplifies in the limit. From the $F_{(r)(t)}$ component of the field strength tensor one can obtain
\prov{A_\tau&=-B\chi\(1-\frac{2\(1+B^2a^2\)^2}{\(1+B^4a^4\)\(1+\cos^2\vartheta\)+2B^2a^2\sin^2\vartheta}\)\rvc\\A_\psi&=\frac{1-B^4a^4}{1+B^4a^4}\frac{2Ba^2\sin^2\vartheta}{\(1+B^4a^4\)\(1+\cos^2\vartheta\)+2B^2a^2\sin^2\vartheta}\rvt}
Again, one can check that quantities evaluated above satisfy the full Einstein--Maxwell system.

\subsection{Near-horizon description of a general MKN black hole}

Although the Ernst--Wild solution is more complicated than the stationary Ernst solution, we have seen that their near-horizon descriptions are equally simple. Let us now turn to the general MKN case, where such simplicity of the near-horizon description is lost. The metric can be written in the form
\rov{\mbs g=\vld f\f(\vartheta\)\(-\frac{\chi^2}{\(Q^2+2a^2\)^2}\bd\tau^2+\frac{\bd\chi^2}{\chi^2}+\bd\vartheta^2\)+\frac{\(Q^2+2a^2\)^2\sin^2\vartheta}{\vld f\f(\vartheta\)}\(\bd\psi-\tilde\omega\chi\bd\tau\)^2\rvc}{mknnhdim}
where the dimensional structural function $\vld f\f(\vartheta\)$ reads
\drov{\vld f\f(\vartheta\)={}&\(1+\frac{1}{4}B^2Q^2\)^2\[Q^2+a^2\(1+\cos^2\vartheta\)\]+\zrov&+2\(1+\frac{1}{4}B^2Q^2\)\[B^2a^2\(Q^2+a^2\)+BQa\sqrt{Q^2+a^2}\]\sin^2\vartheta+\zrov&+B^2\(Ba\sqrt{Q^2+a^2}+Q\)^2\[a^2+\(Q^2+a^2\)\cos^2\vartheta\]\rvt}{mknstrdim}
The dragging constant $\tilde\omega$ can be evaluated using equations of the Harrison transformation in the following form:
\drov{\tilde\omega={}&\frac{1}{\(Q^2+2a^2\)^2}\Bigg\{2BQ^3\(1+\frac{1}{4}B^2Q^2+2B^2a^2\)+4B^3Qa^4+\zrov&+a\sqrt{Q^2+a^2}\[-2\(1-B^4a^4\)+2B^4Q^2a^2+3B^2Q^2+\frac{3}{8}B^4Q^4\]\Bigg\}\rvt}{tommkn}
The $F_{(r)(t)}$ component for a MKN black hole can be straightforwardly calculated (in an algorithmic way utilising a computer algebra system) using defining relations of the Harrison transformation and in the limit it enables us to obtain components of the electromagnetic potential as follows:
\prov{A_\tau=\frac{1}{\vld f\(\vartheta\)}\frac{\chi}{Q^2+2a^2}\[Q\(1-\frac{1}{4}B^2Q^2\)+2Ba\sqrt{Q^2+a^2}\]\vast\{\Bigg[Q\(1-\frac{1}{4}B^2Q^2\)+\nzrov+2Ba\sqrt{Q^2+a^2}\Bigg]^2+\Bigg[\(1+\frac{1}{4}B^2Q^2\)^2a^2+B^2\(Ba\sqrt{Q^2+a^2}+Q\)^2\(Q^2+a^2\)-\lbl{ataumknnh}\zrovn-2\(1+\frac{1}{4}B^2Q^2\)\[B^2a^2\(Q^2+a^2\)+BQa\sqrt{Q^2+a^2}\]\Bigg]\sin^2\vartheta\vast\}\rvc\\A_\psi=-\frac{\tilde\omega}{2\vld f\f(\vartheta\)}\frac{(Q^2+2a^2)^2\[Q\(1-\frac{1}{4}B^2Q^2\)+2Ba\sqrt{Q^2+a^2}\]\sin^2\vartheta}{1+\frac{3}{2}B^2Q^2+2B^3Qa\sqrt{Q^2+a^2}+B^4\(\frac{1}{16}Q^4+Q^2a^2+a^4\)}\lbl{apsimknnh}\rvt}
Even in this most complicated case, $A_\psi$ contains the overall $\sin^2\vartheta$ factor.

\section{Near-horizon degeneracy of extremal MKN class}

\begin{textblock}{}(96.2,284.8)
{\fontfamily{ppl}\selectfont
 53}
\end{textblock}
\vspace{0.25em}

It is carefully discussed e.g. by \cite{BiKaLe} that external stationary, axisymmetric magnetic fields are expelled from the degenerate horizons. This is referred to as a black-hole Meissner effect. However, that discussion focuses on the test fields. It is natural to ask whether such effect is present even in the strong field regime, i.e. for a general MKN black hole. The question is rather difficult because of the non-linear effects (like the presence of the frame dragging in the stationary Ernst spacetime, which is obtained from originally static Reissner--Nordström solution). The presence of the Meissner effect in the MKN setup has been demonstrated using magnetic fluxes by \cite{KarVok91, KarBud}. However, the discussion is quite involved. We will show that there is another indication of the Meissner effect in the strong field regime for a general MKN black hole.

To illustrate our point, we begin with a simple special case. It is characterised by specific relationship among the parameters entering the solutions:
\rov{B=\frac{-2Q\sqrt{Q^2+a^2}\pm2\(Q^2+2a^2\)}{a\(3Q^2+4a^2\)}\rvt}{brbeq}
Under this constraint the degenerate horizon becomes spherically symmetrical and the near-horizon limit is just
the Robinson--Bertotti solution, i.e. \eqref{knnh} with $a=0$, which involves no magnetic field (see \cite{dipl}).

Leaving the special cases aside, we wish to point out that the metrics \eqref{knnh} and \eqref{mknnhdim} are mathematically equivalent in general. After analysing the scaling of the Killing vectors, we can deduce that it is possible to describe the near-horizon limit of extremal black holes in strong magnetic fields by some effective parameters $\hat M,\hat a,\hat Q$ instead of the parameters $M, a, Q, B$ that we used so far.

To be more specific, we should identify the coefficients inside function $\vld f\f(\vartheta\)$ in \eqref{mknstrdim} in accordance with the \emph{form} of the near-horizon limiting metric of Kerr--Newman solution \eqref{knnh} which is
\rov{\vld f\f(\vartheta\)=\hat M^2+\hat a^2\cos^2\vartheta\rvt}{}
If we look at the particular form of 
 function \eqref{mknstrdim}, we see that we have to pair up all the sine squared terms with cosine squared terms. The resulting constant part will be the effective mass squared
\drov{\hat M^2=\(1+\frac{1}{4}B^2Q^2\)^2\(Q^2+a^2\)+B^2\(Ba\sqrt{Q^2+a^2}+Q\)^2a^2+\zrov+2\(1+\frac{1}{4}B^2Q^2\)\[B^2a^2\(Q^2+a^2\)+BQa\sqrt{Q^2+a^2}\]\rvc}{mqeff}
whereas the remaining coefficient in front of the cosine squared will be the effective rotation parameter squared
\drov{\hat a^2=\(1+\frac{1}{4}B^2Q^2\)^2a^2+B^2\(Ba\sqrt{Q^2+a^2}+Q\)^2\(Q^2+a^2\)-\zrov-2\(1+\frac{1}{4}B^2Q^2\)\[B^2a^2\(Q^2+a^2\)+BQa\sqrt{Q^2+a^2}\]\rvt}{aqeff}
Both these quantities are non-negative, which can be most easily manifested by simplifying their square roots to following forms\footnote{We may also ask whether $\hat M$ itself is non-negative. Solving quadratic equation $\hat M=0$ with respect to $B$, one readily finds that it has no real roots.}:
\prov{\hat M&=\sqrt{Q^2+a^2}\(1+\frac{1}{4}B^2Q^2+B^2a^2\)+BQa\rvc\\\hat a&=a\(1-\frac{3}{4}B^2Q^2-B^2a^2\)-BQ\sqrt{Q^2+a^2}\rvt}
Being led by the Kerr--Newman solution, we can define effective charge by $\hat Q^2=\hat M^2-\hat a^2$, which gives
\rov{\hat Q=Q\(1-\frac{1}{4}B^2Q^2\)+2Ba\sqrt{Q^2+a^2}\rvt}{qeff}
Now one should simply substitute the expressions for $\hat Q, \hat a$ in place of $Q, a$ into the metric \eqref{knnh} and it turns out that the result is the metric \eqref{mknnhdim} expressed in rescaled coordinates (due to different scaling of the Killing vectors). Then we should do the same for the electromagnetic potential, i.e. insert the expressions for $\hat Q, \hat a$ in place of $Q, a$ in \eqref{aknnh} and compare with \eqref{ataumknnh} and \eqref{apsimknnh}, again with coordinates rescaled the same way. We successfully performed this calculation using computer algebra system and verified that the description of the near-horizon limit of MKN black hole using the effective parameters indeed works. Therefore the solution \eqref{mknnhdim} is mathematically equivalent to the solution \eqref{knnh} as we mentioned above.

Let us make a final point: there is nothing like an effective parameter $\hat B$; the parameter $B$ has been absorbed into the structure of $\hat Q$ and $\hat a$. We have already noted that the azimuthal component of the electromagnetic potential \eqref{apsimknnh} for the near-horizon description of the MKN black hole is proportional to $\tilde\omega$. Having established the equivalence above, we can be more explicit and say that $A_\psi$ is proportional to $\hat a$ (as $\tilde\omega$ is) and that it has the same \emph{internal} nature as the azimuthal component of the electromagnetic potential for the Kerr--Newman solution. The external field does not appear in the near-horizon limit. This is the indication of the presence of the Meissner effect in the MKN class of black holes.

\begin{textblock}{}(96.2,284.8)
{\fontfamily{ppl}\selectfont
 54}
\end{textblock}
\vspace{0.25em}

\acknowledgement
{The work was supported by research grants: GAUK No. 606412, SVV No. 265301 and GAČR No. 14-37086 G.}

\begin{textblock}{}(96.2,284.8)
{\fontfamily{ppl}\selectfont
 55}
\end{textblock}

\end{article}


\begin{thebibliography}{}

\bibitem[\it Bardeen and Horowitz(1999)]{BardHorow}
  {  Bardeen}, J., {  Horowitz}, G. T.
  \emph{Extreme Kerr throat geometry: A vacuum analog of $\mrm{AdS_2\times S^2}$}.
  Physical Review D, {\bf 60}, 104030,
  1999, 10 pages (arXiv:hep-th/9905099).

\bibitem[\it Bičák and Janiš(1985)]{BiJa85}
  {  Bičák}, J., {  Janiš}, V.
  \emph{Magnetic fluxes across black holes}.
  Monthly Notices of the Royal Astronomical Society, {\bf 212}, 899--915,
  1985.

\bibitem[\it Bičák et al.(2007)]{BiKaLe}
  {  Bičák}, J., {  Karas}, V., {  Ledvinka}, T.
  \emph{Black holes and magnetic fields} in \emph{Black Holes from Stars to Galaxies --- Across the Range of Masses}. Edited by V. Karas and G. Matt. Proceedings of IAU Symposium No. 238, 139--144.
  Cambridge University Press, 2007 (arXiv:astro-ph/0610841).

\bibitem[\it Baez and Díaz(1986)]{DiazBaez86}
  {  Bretón Baez}, N., {  García Díaz}, A.
  \emph{Magnetic generalizations of the Reissner--Nordstrøm class of metrics and Bertotti--Robinson solutions}.
  Il Nuovo Cimento B, {\bf 91}, 83--99,
  1986.

\bibitem[\it Carter (1972)]{Carter73}
  {  Carter}, B.
  \emph{Black hole equilibrium states}.
  General Relativity and Gravitation, {\bf 41}, 2873--2938,
  2009.
  General Relativity and Gravitation, {\bf 42}, 653--744,
  2010.
  Original edition of both parts in
  \emph{Black holes (Les Houches lectures)},
  eds. B. DeWitt and C. DeWitt,
  Gordon and Breach, New York, 1972.

\bibitem[\it Ernst(1976)]{Ernst76}
  {  Ernst}, F. J.
  \emph{Black holes in a magnetic universe}.
  Journal of Mathematical Physics, {\bf 17}, 54--56, 1976.
  
\bibitem[\it Ernst and Wild(1976)]{ErnstWild}
  {  Ernst}, F. J., {  Wild}, W. J.
  \emph{Kerr black holes in a magnetic universe}.
  Journal of Mathematical Physics, {\bf 17}, 182--184, 1976.
  
\bibitem[\it Frolov and Novikov(1998)]{FrolNov}
  {  Frolov}, V. P., {  Novikov}, I. D.
  \emph{Black hole physics: basic concepts and new developments}.
  Springer, 1998.

\bibitem[\it Gibbons et al.(2013)]{Pope}
  {  Gibbons}, G. W., {  Mujtaba}, A. H., {  Pope}, C. N.
  \emph{Ergoregions in Magnetised Black Hole Spacetimes}.
  Classical and Quantum Gravity, {\bf 30}, 125008,
  2013, 23 pages (arXiv:1301.3927 [gr-qc]).
  
\bibitem[\it Hejda(2013)]{dipl}
  {  Hejda}, F. \emph{Particles and fields in curved spacetimes (selected problems)}.
  Master thesis.
  Prague, 2013.

\bibitem[\it Karas and Budínová(2000)]{KarBud}
  {  Karas}, V., {  Budínová}, Z.
  \emph{Magnetic Fluxes Across Black Holes in a Strong Magnetic Field Regime}.
  Physica Scripta, {\bf 61}, 253--256,
  2000.

\bibitem[\it Karas and Vokrouhlický(1991)]{KarVok91}
  {  Karas}, V., {  Vokrouhlický}, D.
  \emph{On interpretation of the magnetized Kerr--Newman black hole}.
  Journal of Mathematical Physics, {\bf 32}, 714--716, 1991.

\bibitem[\it Melvin(1964)]{Melvin64}
  {  Melvin}, M. A.
  \emph{Pure magnetic and electric geons}.
  Physics Letters, {\bf 8}, 65--68,
  1964.

\bibitem[\it Stuchlík and Hledík(1999)]{StuchHle99}
  {  Stuchlík}, Z., {  Hledík}, S.
  \emph{Photon capture cones and embedding diagrams of the Ernst spacetime}.
  Classical and Quantum Gravity, {\bf 16}, 1377--1387,
  1999 (arXiv:0803.2536 [gr-qc]).

\end{thebibliography}
\end{document}